\documentclass[utf8,numbers]{FrontiersinHarvard}
\usepackage{url,hyperref,lineno,microtype,subcaption}
\usepackage[onehalfspacing]{setspace}

\usepackage{amsmath,amssymb,amsfonts}
\usepackage{algorithm}
\usepackage{algorithmicx}
\usepackage{algpseudocode}
\usepackage{setspace}
\usepackage{bm}
\usepackage{graphicx}
\usepackage{textcomp}
\usepackage[most]{tcolorbox}
\usepackage{xcolor}
\usepackage{tikz}
\usetikzlibrary{spy}
\usepackage{pgfplots}
\pgfplotsset{compat=1.18}
\usetikzlibrary{quantikz2}
\usepackage{float}
\usepackage{subcaption}
\usepackage[T1]{fontenc}
\usepackage{verbatim}
\usepackage{hyperref}
\usepackage{makecell}
\usepackage{booktabs}
\usepackage{array} 
\usepackage{amsthm}
\usepackage{graphbox}
\usepackage{adjustbox}
\usepackage{balance}
\usepackage{multirow}
\usepackage[capitalize]{cleveref}
\usepackage{footnote}
\usepackage{bm}
\usepackage{threeparttable}
\usepackage{caption}
\usepackage{lipsum}
\usepackage{physics}
\usepackage{braket}
\usepackage{stfloats}
\usepackage{placeins}
\usepackage{cuted}
\usepackage{multicol}
\usepackage[top=0.71in, bottom=1in, left=0.7in, right=0.7in]{geometry}
\def\BibTeX{{\rm B\kern-.05em{\sc i\kern-.025em b}\kern-.08em
    T\kern-.1667em\lower.7ex\hbox{E}\kern-.125emX}}

\theoremstyle{remark}

\theoremstyle{definition}

\linenumbers

\def\keyFont{\fontsize{8}{11}\helveticabold }
\def\firstAuthorLast{A. S. Cacciapuoti {et~al.}}
\def\Authors{Angela~Sara~Cacciapuoti, Claudio Pellitteri, Marcello~Caleffi}
\def\Address{The authors are with the \href{www.quantuminternet.it}{www.QuantumInternet.it} research group, University of Naples Federico II, Naples, 80125 Italy.}
\def\corrAuthor{Angela Sara Cacciapuoti}

\def\corrEmail{angelasara.cacciapuoti@unina.it}

\begin{document}
\onecolumn
\firstpage{1}

\title[Spatial Path Superposition: A No-Cloning Perspective on MIMO Diversity]{Spatial Path Superposition: A No-Cloning Perspective on MIMO-Inspired Spatial Diversity} 

\author[\firstAuthorLast ]{\Authors}
\extraAuth{}
\address{}
\correspondance{}

\maketitle
\begin{abstract}
Spatial diversity is a key mechanism in classical Multi-Input Multi-Output (MIMO) communications, where multiple propagation alternatives provide redundant observations that improve reliability in noisy environments. A direct translation of this principle to quantum communications is forbidden by the no-cloning theorem, since an unknown quantum state cannot be copied and transmitted through independent links. In this Perspective article, we argue that spatial path superposition provides a no-cloning-compatible analogue of the spatial-diversity principle of MIMO. Rather than duplicating the transmitted quantum state, a quantum carrier is coherently delocalized over multiple communication links, so that alternative spatial evolutions are experienced in quantum superposition. In this framework, classical redundancy is replaced by spatial coherence, and the communication link itself becomes part of the quantum dynamics. Indeed, spatial coherence can enhance the fidelity of the received quantum state, providing the quantum counterpart of the classical diversity gain, and can be converted into bipartite and multipartite entanglement during propagation. This implies that, under suitable coherent path control, noisy channel evolutions can contribute constructively to entanglement generation.
This Perspective identifies spatial path superposition as a quantum-native primitive for the Quantum Internet, linking no-cloning-compatible spatial diversity, fidelity enhancement, entanglement generation, and coherent path control.
\tiny
 \keyFont{\section{Keywords:} Quantum MIMO, Spatial Path Superposition, Entanglement, Spatial Diversity, Quantum Internet}
\end{abstract}

\section{Introduction}
\label{sec:01}
Classical Multi-Input Multi-Output (MIMO) communication exploits the spatial dimension of the wireless channel through multiple antennas at the transmitter and receiver, thereby improving reliability, increasing throughput, and supporting multiuser multiplexing. Among the different mechanisms enabled by MIMO, spatial diversity plays a particularly relevant role: the same information is conveyed through multiple spatially distinct propagation paths, so that the resulting redundant observations can be combined at the receiver to reduce the probability of error \cite{Ala-98}. This principle has become one of the key ingredients behind the robustness of classical wireless communication systems.

Transferring this intuition to the quantum domain is fundamentally non-trivial \cite{CalCac-26,CacCalMet-20,CacCal-26}. A direct quantum counterpart of classical spatial diversity would require sending multiple identical copies of an unknown quantum state through different communication links. This strategy is forbidden by the no-cloning theorem \cite{WooZur-82}. Therefore, the classical MIMO intuition cannot be imported into quantum communications simply by replacing classical signals with quantum states. 

The relevant question is instead whether the state encoded in a quantum carrier can be conveyed by exploiting multiple communication links without requiring copies of that state.

Our perspective is that exploiting quantum coherence at the propagation level provides a natural answer to this question.
Specifically, rather than duplicating the transmitted quantum state, a single quantum carrier, such as a photon, can be coherently delocalized over multiple spatial trajectories \cite{ChiKri-19,CalCac-20,CalSimCac-23,PelCalCac-25,SkPelCal-26,PelSkCal-26}. 
This regime corresponds to a quantization of the propagation process itself: the communication link is no longer regarded only as a fixed classical structure through which a quantum state propagates, but becomes part of the quantum dynamics via the channel map associated with each link.

This observation suggests a no-cloning-compatible analogue of MIMO spatial diversity. 
The analogy is deliberately restricted to the diversity principle: it does not aim to reproduce the full MIMO toolbox. Instead, spatial path superposition replaces classical redundancy with quantum coherence. Multiple propagation alternatives are exploited simultaneously, not by creating redundant copies of a quantum state, but by coherently controlling the path degree-of-freedom of a quantum carrier. 
The resulting mechanism relies on interference patterns activated by alternative channel evolutions experienced by the state, thereby providing a \textit{MIMO-inspired quantum spatial-diversity principle} without violating the no-cloning theorem.

The remainder of the paper is organized as follows. 
\cref{sec:02} develops the interpretation of spatial path superposition as a no-cloning-compatible analogue of MIMO spatial diversity. 
\cref{sec:03} reviews the formal framework of spatially superposed communication links and discusses its relation with other coherent-control frameworks, including discrete-time quantum walks and the quantum switch. 
\cref{sec:04} discusses how path coherence can be converted into entanglement during propagation. 
\cref{sec:05} discusses the scope of this Perspective and the networking implications of spatial path superposition.

\section{Spatial Path Superposition as a No-Cloning-Compatible Spatial Diversity}
\label{sec:02} 

Classical spatial diversity exploits redundant observations of the same copyable information transmitted over multiple independently affected links. In the quantum domain, the direct analogue of this strategy is forbidden by the no-cloning theorem. The quantum alternative is not to duplicate the state, but to exploit path coherence: a quantum carrier is not constrained to propagate on a single, well-defined link, but can be coherently delocalized over multiple spatially distinct links. 

As a result, not only the internal degrees of freedom of the carrier, but also the very same instantiation of the communication channels through which it propagates, can acquire a genuinely quantum character. Within this framework, communication links are no longer regarded only as fixed classical structures, but can instead be placed in coherent superposition, with a control degree-of-freedom encoding the path information. 

This provides an operational and \textit{no-cloning-compatible analogue of MIMO spatial diversity}: multiple propagation alternatives are exploited simultaneously, but the unknown quantum state is never duplicated.

This mechanism should be understood as replacing classical redundancy with quantum coherence: in spatial path superposition the receiver exploits interference between alternative channel evolutions experienced by the same quantum carrier. 
The broad framework of coherently controlled quantum trajectories has been shown to provide significant communication advantages, by outperforming strategies, in which the propagation process is not quantized, even in regimes where individual channels are severely noisy, including zero-quantum-capacity regimes \cite{ChiKri-19,CalCac-20,CalSimCac-23,PelCalCac-25,SkPelCal-26,PelSkCal-26,ChaShiDur-2025}.

\begin{figure*}[t]
    \centering
    \begin{minipage}[t]{0.48\textwidth}
        \centering
        \includegraphics[width=\linewidth,height=4cm]{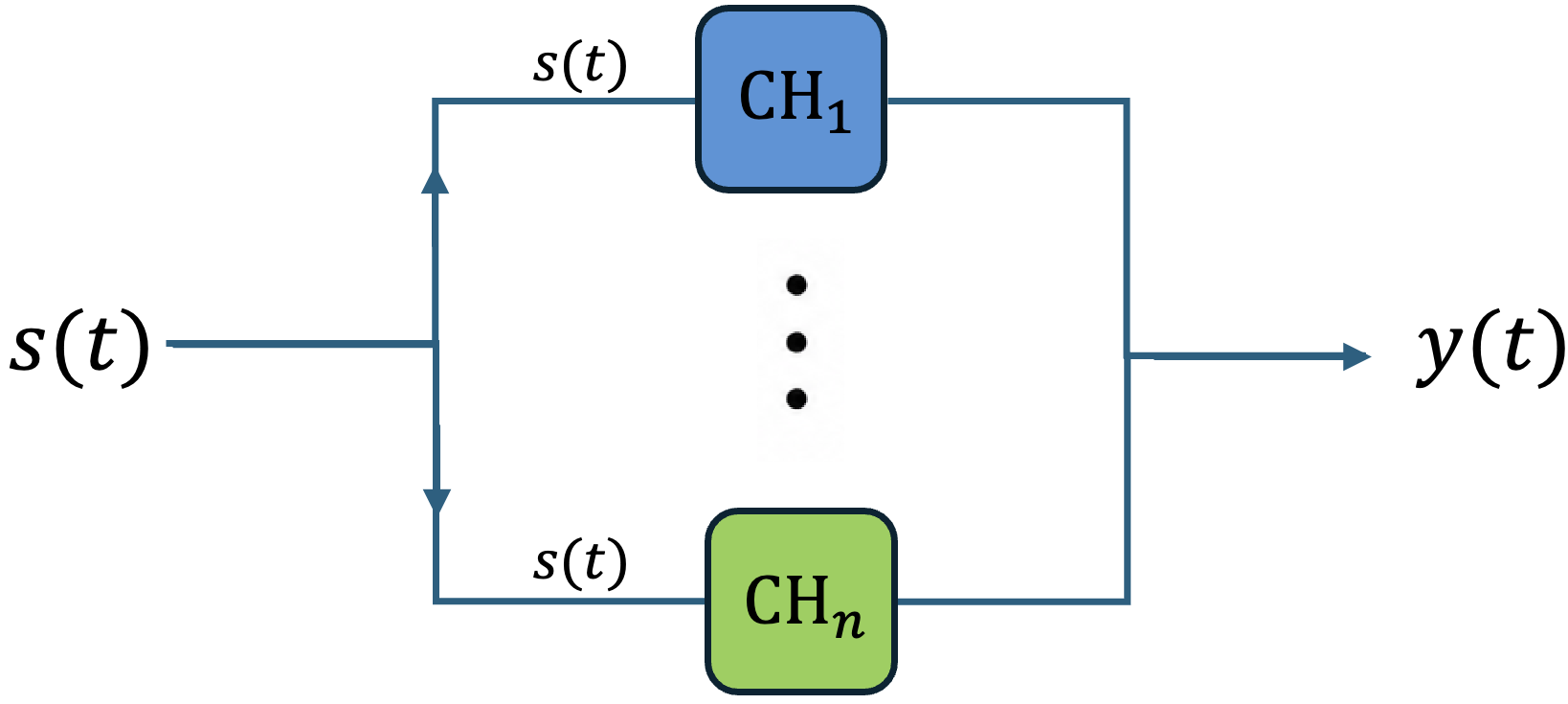}
        \small (a) Classical Spatial Diversity
    \end{minipage}
    \hfill
    \begin{minipage}[t]{0.48\textwidth}
        \centering
        \includegraphics[width=\linewidth,height=4cm]{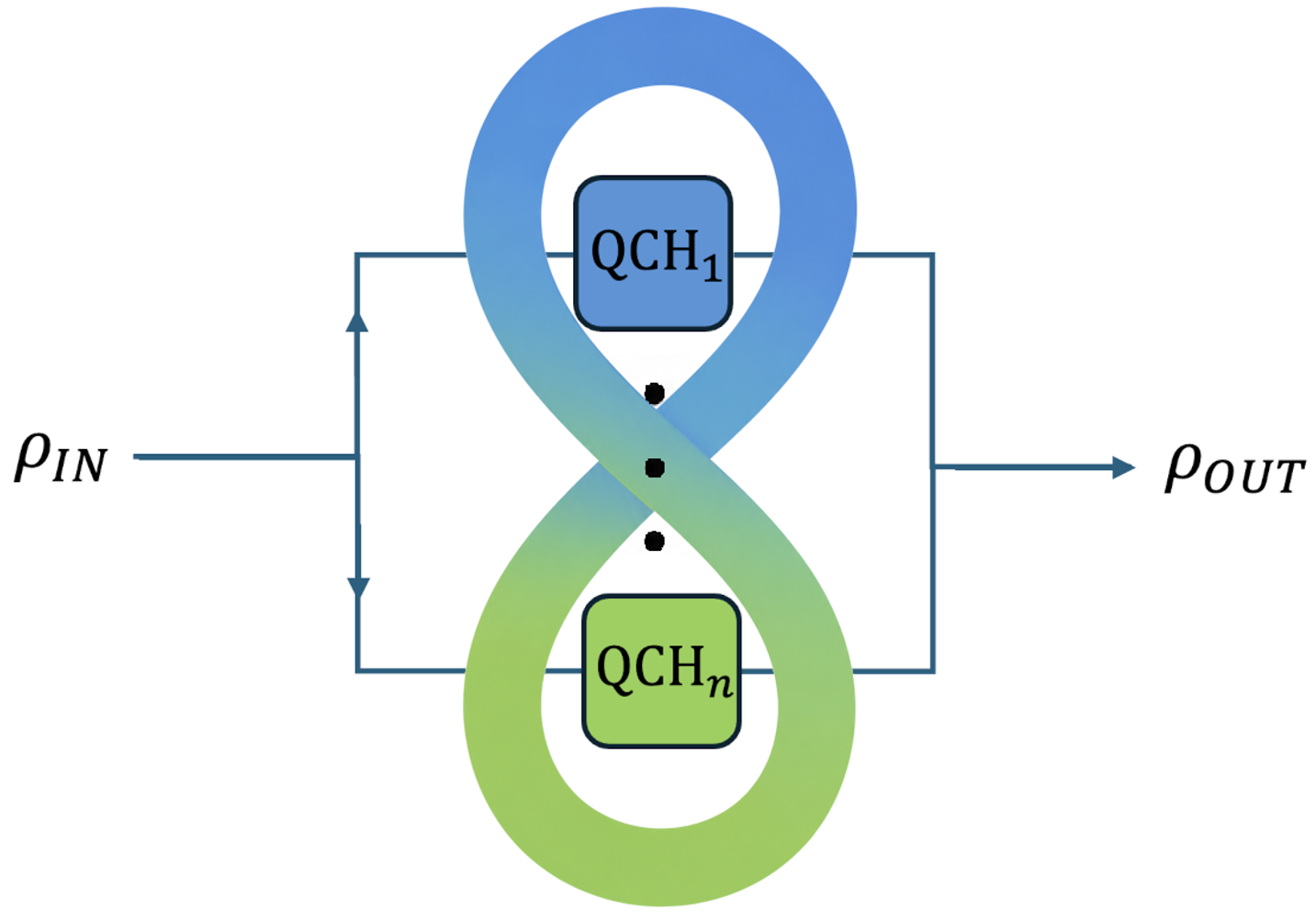}
        \small (b) Spatial Path Superposition 
    \end{minipage}
    \caption{Pictorial representations of classical MIMO Spatial Diversity and Spatial Path Superposition. In the classical case, redundant observations of the same information are obtained through multiple spatial links. In the quantum case, a single carrier is coherently delocalized over multiple link alternatives, without duplicating the transmitted quantum state.}
    \label{fig:01}
\end{figure*}

Figure~\ref{fig:01} provides a pictorial comparison between classical MIMO spatial diversity and spatial path superposition, while Table~\ref{tab:01} summarizes their main conceptual differences.

In the above perspective, the operational counterpart of the classical diversity gain is the improvement of the transmitted quantum state quality, typically quantified through output fidelity \cite{Sch-96}. 
In classical MIMO spatial diversity, redundant observations are combined to reduce the probability of error. In spatial path superposition, alternative channel evolutions interfere coherently, and this interference can increase the fidelity of the received state with respect to the target state. 
In this sense, fidelity improvement plays the role of a quantum analogue of diversity gain: it quantifies the benefit of exploiting multiple propagation alternatives without copying the unknown quantum state.
This advantage is intrinsically coherence-based. 

In this sense, the quantum diversity gain can be directly quantified by the fidelity improvement achieved through spatial path superposition with respect to the best individual path, since fidelity measures how closely the resulting state approaches the desired target state.

This is a genuinely quantum form of diversity that is inaccessible to strategies based on approximate clones of the transmitted state, whose fidelity is necessarily limited by the no-cloning theorem.

\begin{table*}[t]
\caption{Comparison between MIMO spatial diversity and spatial paths superposition.}
\centering
\renewcommand{\arraystretch}{1.5}
\begin{tabular}{p{4cm}p{5cm}p{5cm}}
\hline\hline
\textbf{Feature} & \textbf{Classical MIMO spatial diversity} & \textbf{Spatial Path Superposition} \\
\hline
Information resource & Classical Information & Quantum state carried by a quantum carrier \\
\hline
Parallelism & Redundant observations over multiple links & Coherent superposition of link alternatives\\
\hline
Copying & Allowed for classical information & Forbidden for unknown quantum states \\
\hline
Robustness mechanism & Redundancy and combining & Interference between channel evolutions \\
\hline
Operational gain & Reduced error probability & Enhanced output fidelity \\
\hline
Control variable & Antenna/path weights & Path-control degree-of-freedom \\
\hline
Network Interpretation&
Path diversity & No-cloning-compatible route diversity\\
\hline\hline
\end{tabular}
\label{tab:01}
\end{table*}
\section{Spatial Path Superposition: Formalism and Related Frameworks}
\label{sec:03}
This section summarizes the minimal formal framework needed to describe spatially superposed communication links and to distinguish them from related frameworks based on the coherent control of quantum trajectories.

\subsection{Mathematical Formalism}

The superposition of spatially distinct communication links is based on an extension of the Hilbert space associated with the quantum carrier \cite{ChiKri-19}, by introducing the \emph{vacuum state}, which accounts for the possibility that a given link is not traversed by the carrier. Together with a quantum control system, typically a qubit or, more generally, a qudit, the vacuum extension allows one to describe a carrier propagating through either a single link or a coherent superposition of different links. The control system encodes the path information.

For clarity, we focus on the case of two links. However, the same construction extends naturally to the coherent superposition of $n$ links. Let $\mathcal{F}(\cdot)$ and $\mathcal{N}(\cdot)$ be two quantum channel maps, with Kraus decompositions $\{F_i\}$ and $\{N_j\}$, respectively, describing the quantum evolutions associated with two alternative communication links, i.e., $
\mathcal{F}(\rho)=\sum_i F_i\rho F_i^\dagger$,
$\mathcal{N}(\rho)=\sum_j N_j\rho N_j^\dagger$,
with $\sum_i F_i^\dagger F_i=\sum_j N_j^\dagger N_j=I$. The coherent superposition of these links, in which the alternatives are coherently controlled by a quantum degree of freedom rather than classically selected allowing the delocalization of the carrier on the links, is described by the map:
\begin{equation}
\label{eq:channel_superposition_general}
\mathcal{S}(\mathcal{F},\mathcal{N})(\rho_t)
=
\sum_{ij}
S_{ij}(\rho_t\otimes\rho_c)S^\dagger_{ij},
\end{equation}
where $\rho_t$ is the input state and $\rho_c$ is the state of the control system. The Kraus operators of the extended map are:
\begin{equation}
\label{eq:Sij_general}
S_{ij}
=
\beta_j F_i \otimes \ket{0_c}\bra{0_c}
+
\alpha_i N_j\otimes \ket{1_c}\bra{1_c},
\end{equation}
where, the coefficients $\{\alpha_i\}$ and $\{\beta_j\}$ are the vacuum amplitudes associated with the two channels and satisfy the normalization conditions
\begin{equation}
\sum_i |\alpha_i|^2 = 1,
\qquad
\sum_j |\beta_j|^2 = 1.
\end{equation}

These amplitudes are not fixed by the reduced channel maps alone. 
They depend on the underlying system-environment interaction, so different physical implementations of the same channel may lead to different vacuum extensions and, therefore, to different interference effects under coherent superposition \cite{ChiKri-19,PaLuNo-23, SkPelCal-26,PelSkCal-26}. 
For instance, in the interferometric photonic implementation reported in \cite{Pang-23}, the control degree of freedom is encoded in the path taken by the carrier, while the target system is encoded in polarization. Each channel realization is implemented through liquid-crystal wave-plates that reproduce the relevant Pauli operations via applied voltages, and a phase control plate sets the relative phase between the superposed paths. In such a setup, the effective vacuum amplitudes are controlled through the calibrated wave-plate voltages, phase shifters, and, when required, amplitude-balancing elements such as variable beam splitters or attenuators — although the specific mapping between these hardware parameters and the vacuum amplitudes remains implementation-dependent, and must be derived and calibrated for the particular physical realization.
This dependence is key for the fidelity and entanglement-generation effects discussed here.
For a fixed physical realization, these vacuum amplitudes are treated here as fixed complex coefficients, rather than as additional dynamical variables explicitly parameterized by propagation distance or time; any distance or time dependence would instead enter through the underlying channel realization and its vacuum extension.

\subsection{Relation with other coherent-control frameworks}

Spatial path superposition belongs to the broader framework of coherent control of quantum trajectories \cite{ChiKri-19}. Within this framework, superposition can arise in two fundamentally distinct ways. The first, considered in this Perspective, is the superposition of spatial paths, where the carrier coherently propagates through different physical links. The second is the superposition of causal orders, implemented by the quantum switch, where the order in which two or more channels are applied is coherently controlled.  
Formally, for two quantum channel maps $\mathcal{F}(\cdot)$ and $\mathcal{N}(\cdot)$ with Kraus operators $\{F_i\}$ and $\{N_j\}$, respectively, the quantum switch coherently combines the two causal orders $\mathcal{F}\circ\mathcal{N}$ and $\mathcal{N}\circ\mathcal{F}$. And the corresponding Kraus operators of the extended map are $ S_{ij}=F_i N_j \otimes \ket{0}\bra{0} + N_j F_i \otimes \ket{1}\bra{1}$.

From the perspective developed in this paper, the quantum switch can also be regarded as a form of coherence-assisted diversity, because the communication process benefits from controlling multiple order alternatives, without improving the individual channels themselves \cite{CalCac-20,EbSalChi-18,CalSimCac-23}. Moreover, as shown in \cite{PelCalCac-25}, the quantum switch can be seen as a two-step application of spatial path superposition. 

Nevertheless, spatial path superposition and causal-order superposition are operationally distinct. The quantum switch controls the order of channel applications, whereas the spatial path superposition framework controls the physical link traversed by the carrier.
Although the quantum switch offers deep insights into the foundational structure of the quantum theory by challenging the classical notion of a definite temporal order of events, its experimental realization remains highly demanding \cite{WeiTisZha-19,RubRozWal-17}. Indeed, its physical interpretation is still under active debate, particularly regarding whether current implementations genuinely exhibit indefinite-causal order or rather simulate it \cite{WeiTisZha-19,RubRozWal-17,HamKabBru-25}.
Differently, spatial path superposition provides a more experimentally accessible and conceptually transparent approach \cite{AbbWecBra-20}. Spatial path superposition can be implemented using well-established techniques such as interferometric setups \cite{PaLuNo-23} or quantum-walk-inspired architectures \cite{VenAnd-2012,Kempe2003}.

Discrete-time quantum walk (DTQW) provides a useful example of the above consideration. In DTQW, the coin degree-of-freedom plays the role of a control system, while the position degree-of-freedom encodes the spatial alternatives explored by the carrier. The walker evolves through interference among multiple paths, producing non-classical propagation features \cite{VenAnd-2012,Kempe2003}. 
Thus, DTQW can be viewed as specific instantiation of the broader spatial path-superposition framework \cite{PelSkCal-26,PelCalCac-25}. 

\section{Entanglement Generation as an Operational Consequence of Path Coherence}
\label{sec:04}
Spatial path superposition, introduced above as a no-cloning-compatible form of quantum spatial diversity, is more than a robustness mechanism. 
By generating interference terms between alternative channel evolutions, path coherence can act as an active networking resource and produce non-classical correlations during propagation. 
Thus, propagation is no longer a passive process through which a quantum state is merely transported: under coherent control, it can become part of the mechanism that generates entanglement.

This perspective is particularly relevant for quantum networks, where entanglement is the key communication resource. The generated correlations do not require a priori distribution of entanglement: they arise as a direct operational consequence of coherently controlling the path degree-of-freedom. When a carrier is delocalized over multiple communication links, the path-control degree-of-freedom and the target degree-of-freedom are jointly involved in the channel evolution. After measuring the control system, path coherence can be converted into entanglement among the target systems, starting from initially separable inputs. Thus, the same mechanism that replaces classical redundancy in the quantum-diversity analogy can also transform propagation itself into a resource for entanglement distribution.

Recent results make this connection explicit \cite{SkPelCal-26, PelSkCal-26}.
Starting from initially separable input states, noisy communication channels placed in spatial superposition can generate Bell, GHZ, and W-type entangled states during the distribution process. 
Even depolarizing channels, usually regarded as a worst-case noise model, can contribute to Bell-state generation in the bipartite case and to GHZ-state generation in multipartite extensions. 
W-state generation requires a more structured superposition of noisy operations, reflecting the different entanglement structure of W states.

These results also suggest a different role for noise in quantum networking. Under coherent path control, noisy channel evolutions can contribute to entanglement generation through interference \cite{PelSkCal-26}. This constructive effect depends on the channel structure and on its physical vacuum extension, including the associated amplitudes and phases, and it requires the preservation of control coherence. Strictly speaking, it is therefore not the noise that is constructive, but the engineered coherence: the effect should be understood as an engineering possibility that emerges when the amplitudes and phases of a coherently controlled propagation process can be properly tuned, rather than as a generic beneficial effect of noise.

\section{Discussion}
\label{sec:05}
The analogy developed in this Perspective is deliberately restricted to the spatial-diversity principle, not to the full MIMO framework. The correspondence is between classical path diversity and spatial path coherence: both exploit multiple propagation alternatives, but classical diversity relies on redundant observations of copyable information, whereas spatial path superposition relies on interference between alternative evolutions of a quantum carrier.

This perspective also clarifies the assumptions under which spatial path superposition becomes an operational resource. 
The control system governs the quantum coherence between the alternative communication links. 
If the control decoheres, the interference terms responsible for fidelity enhancement and entanglement generation are suppressed, and the process reduces to an effective classical mixture. Nevertheless, treating the control system as effectively protected is well motivated. 
From a theoretical standpoint, the most demanding noise is assigned to the target system, namely the subsystem carrying the quantum information or correlations to be transmitted or generated. By contrast, the control system is typically low-dimensional and can in principle be protected more easily, for instance through decoherence-free subspaces. 
From a physical standpoint, in several implementations the degree-of-freedom encoding the control does not undergo the same noisy evolution as the target. For example, in interferometric realizations, the path degree-of-freedom can encode the control while the noisy channel acts on an internal degree-of-freedom of the carrier \cite{PaLuNo-23,AlpHadWen-04,LidChuWha-98}. Moreover, moderate noise on the control system has been shown not to qualitatively destroy the operational advantages of the spatial superpositions framework \cite{ChaShiDur-2025}. 

Experimental implementations nevertheless require path coherence and relative-phase stability among the alternative trajectories. In interferometric realizations, the relative phase between the branches must be actively set and stabilized, and the different interference regimes are selected precisely by acting on such a phase \cite{PaLuNo-23}.

In this sense, fidelity enhancement or entanglement generation should be interpreted as an engineering possibility enabled by coherent control, rather than as a speculative effect.

Spatial path superposition should not be regarded as a general replacement for conventional direct-transmission or repeater-based entanglement-distribution schemes, but rather as a complementary coherence-enabled primitive. Its distinctive advantage arises in regimes where multiple propagation alternatives can be coherently controlled, allowing their interference to enhance output fidelity or generate entanglement during propagation. Conversely, when such path coherence cannot be established or maintained, the distinctive interference-based mechanism of spatial path superposition is no longer available. A quantitative resource-normalized comparison with conventional entanglement-distribution architectures is beyond the scope of the present Perspective and remains an important direction for future research. 

At the networking level, an important open problem is whether coherent path control can be lifted from a link-level resource to a full quantum-native routing primitive for multi-hop quantum networks. Although the formal construction extends to multiple coherently superposed links, its physical scaling requires increasing path-control overhead, together with synchronization and control in the presence of loss, relative-phase errors, and hardware non-idealities.

Overall, the proposed viewpoint reframes spatial diversity under the constraints of quantum mechanics. Within this view, fidelity improvement plays the role of a quantum diversity gain, while entanglement generation shows that path coherence can become an active resource for quantum networking.

\section*{Funding}
This work has been funded by the European Union under Horizon Europe ERC-CoG grant QNattyNet, n.101169850. Views and opinions expressed are however those of the author(s) only and do not necessarily reflect those of the European Union or the European Research Council Executive Agency. Neither the European Union nor the granting authority can be held responsible for them.

\bibliographystyle{IEEEtran}
\bibliography{Bibliography.bib}

\end{document}